\documentclass[11pt]{article}
\usepackage[utf8]{inputenc}
\usepackage[left=1in,right=1in]{geometry}
\usepackage{authblk}
\usepackage[pdfstartview=FitH,pdfpagemode=None]{hyperref}
\usepackage{amsmath}
\usepackage{amsfonts}
\usepackage{graphicx,graphbox}
\usepackage{braket}
\usepackage{comment}
\usepackage{cite}
\title{Reparameterization Dependence is Useful for Holographic Complexity}
\author[1,2]{Ayoub Mounim \thanks{E-mail: \texttt{ayoub.mounim@unina.it}}}
\author[1,2]{Wolfgang M\"uck \thanks{E-mail: \texttt{mueck@na.infn.it}}}
\affil[1]{Dipartimento di Fisica ``Ettore Pancini", Universit\`a degli Studi di Napoli ``Federico II" \authorcr Via Cintia, 80126 Napoli, Italy}
\affil[2]{Istituto Nazionale di Fisica Nucleare, Sezione di Napoli \authorcr Via Cintia, 80126 Napoli, Italy}
\date{\today}
\numberwithin{equation}{section}

\newcommand{\ie}{i.e.,\ }
\newcommand{\eg}{e.g.,\ }

\newcommand{\rmd}{\,\mathrm{d}}

\newcommand{\e}[1]{\operatorname{e}^{#1}}

\begin{document}
\maketitle
\begin{abstract}
Holographic complexity in the ``complexity equals action'' approach is reconsidered relaxing the requirement of reparameterization invariance of the action with the prescription that the action vanish in any static, vacuum causal diamond. This implies that vacuum anti-de Sitter space plays the role of the reference state. Moreover, the complexity of an anti-de Sitter-Schwarzschild black hole becomes intrinsically finite and saturates Lloyd's bound after a critical time. It is also argued that several artifacts, such as the unphysical negative-time interval, can be removed by truly considering the bulk dual of the thermofield double state.      
\end{abstract}
\section{Introduction}
\label{intro}

In recent years, interest has been mounting in relating concepts from information theory to quantum field theory and gravity. One aim of these efforts is to shed new light on our understanding of topics such as the nature of space-time, black hole physics and, ultimately, quantum gravity \cite{Ryu:2006ef, Ryu:2006bv, Casini:2011kv}. Holographic complexity \cite{Susskind:2014rva, Susskind:2018pmk} is such a concept where, in the spirit of the AdS/CFT correspondence \cite{Maldacena:1997re, Gubser:1998bc, Witten:1998qj}, a geometrical quantity related to a bulk space-time is identified as a measure of complexity of a certain boundary state. Two competing proposals are the "complexity equals volume" ($C=V$) \cite{Susskind:2014rva, Alishahiha:2015rta} and "complexity equals action" ($C=A$) \cite{Brown:2015bva, Brown:2015lvg} frameworks. In the $C=A$ approach, which we will use exclusively in this paper, complexity is identified with the action evaluated in a bulk region called the Wheeler-de Witt (WdW) patch,\footnote{In the rest of the paper, we will work with the reduced action $I = 16\pi G S$.}
\begin{equation}
\label{intro:complexity}
	\mathcal{C} = \frac{S_{\mathrm{WdW}}}{\pi\hbar}~.
\end{equation}
The WdW patch is defined as the region bounded by the null surfaces anchored at certain times on the space-time boundary (left and right boundaries in the case of two-sided black holes) and, possibly, the black hole singularity. The action on the WdW patch is generically divergent, because of the contribution of the region close to the space-time boundary. Such a divergence is typical in the AdS/CFT correspondence. Several regularization procedures have been studied, and it has been proposed, just as in holographic renormalization \cite{Emparan:1999pm, deHaro:2000vlm, Bianchi:2001kw, Martelli:2002sp, Skenderis:2002wp}, that the divergences can be removed by adding local, covariant counter terms on the boundaries \cite{Carmi:2016wjl, Kim:2017lrw, Akhavan:2019zax, Hashemi:2019xeq}. Alternatively, one can subtract the action of some reference space-time, which is also necessary to calculate the complexity of formation \cite{Chapman:2016hwi}.

Computational complexity is a concept from information theory measuring how difficult it is (or how many steps it takes) to compute (approximately) a desired target state starting from a given reference state using a certain set of elementary operations \cite{Aaronson:2016vto}. The precise definition of computational, or cicruit, complexity depends on the system under consideration, the set of elementary operations, the reference state and a parameter $\epsilon$ that specifies the tolerance with which the target state is reached. Typically, the complexity diverges when $\epsilon \to 0$. A geometric approach to complexity, which can be applied to quantum field theory, was developed in \cite{Nielsen1133}.
In this approach, complexity is evaluated by a weight function evaluated on a trajectory connecting the target and the reference state in some space of unitary operators. Several proposals for the weight function have been investigated in \cite{Jefferson:2017sdb, Chapman:2017rqy, Sinamuli:2019utz, Yang:2017nfn, Khan:2018rzm, Doroudiani:2019llj}.

In this paper, we will reconsider the complexity associated with global AdS space-time and AdS-Schwarzschild black holes. These are the simplest systems and have, of course, been investigated already in the very first papers on holographic complexity \cite{Brown:2015bva, Brown:2015lvg, Lehner:2016vdi, Carmi:2017jqz} and more recently in \cite{Akhavan:2019zax},\footnote{Other settings have been considered, \eg in \cite{Swingle_2018, Cano_2018, Auzzi_2018, Yaraie_2018, Pan_2017, Goto_2019, Balushi:2020wjt, Hashemi:2019aop, Alishahiha:2019cib}.} 
but there are good reasons to have a fresh look at them. If we take global AdS as a reference state, then we would expect that there exists a procedure (not reference subtraction) that yields a vanishing action, reflecting the fact that the complexity of the reference state is, by definition, zero. Adding counter terms does not achieve this and may even leave a logarithmically divergent term \cite{Akhavan:2019zax}. However, if such a procedure can be found, then applying it to the AdS-Schwarzschild black hole would give immediately the complexity of formation without the need of reference subtraction.

We will approach this problem by exploiting the freedom of reparameterization of the null boundaries of the WdW patch. It is known that the minimal action terms required by the variational principle (we will review them in section~\ref{actionWdW}) are not invariant under reparameterization of the null directions on the null boundaries. To achieve invariance, a counter term must be added, which does not interfere with the variational principle \cite{Lehner:2016vdi}. Although this seems inevitable, the necessity of adding such a counter term may be disputed. First, the null boundary terms carry a physical meaning as the heat flux through the boundary \cite{Chakraborty:2019doh}, so that different parameterizations may describe physically different situations. Second, the counter term is not unique \cite{Jubb:2016qzt}.
However, if reparameterization invariance is given up, then one needs a solid, physically motivated criterion for choosing a particular parameterization. Our criterion will be that \emph{the action in any static vacuum causal diamond vanishes}. A causal diamond is defined as a region bounded only by null surfaces. Because global AdS is a static vacuum space-time, this procedure achieves the goal by definition. In the case of the AdS-Schwarzschild black hole, which is also a static vacuum space-time, the region that effectively contributes to the complexity reduces to a region bordering with the black hole singularity. This region lies entirely behind the horizon, which directly incorporates the general expectation that complexity is a probe of the physics behind the horizon. Moreover, because the region near the space-time boundary does not contribute, the resulting complexity is intrinsically finite.

The rest of the paper is organized as follows. In section~\ref{actionWdW}, we briefly review the contributions to the gravitational action on a WdW patch. In section~\ref{ads} we show how, through a parameterization choice, we can set the complexity of empty AdS space-time to zero, effectively making it our pick for the reference state. We then use this idea to compute the complexity of the AdS-Schwarzschild black hole in section~\ref{sads}. The special case of three dimensions, in which the relevant solution is the BTZ black hole, is treated separately in section~\ref{btz}. We conclude in section~\ref{conc}.

\section{Action on a WdW patch}
\label{actionWdW}

The gravitational action on a WdW patch consists of contributions coming form the bulk of the patch, its boundaries and the joints in which the boundaries meet. Given a bulk action, the variational principle imposes a number of boundary and joint terms. The most familiar case for Einstein gravity is the Gibbons-Hawking-York term \cite{York:1972sj, Gibbons:1976ue}, which applies to space-like or time-like boundaries. When the boundary is not smooth, additional joint terms are necessary \cite{Hartle:1981cf, Hayward:1993my}. In the general case, which includes also null boundaries, the boundary and joint terms have been constructed, \eg in \cite{Parattu:2015gga, Parattu:2016trq, Lehner:2016vdi, Jubb:2016qzt}. We refer to \cite{Jubb:2016qzt} for more references to the original work. We shall now review the contributions one at a time. 

The bulk contribution is given by the Einstein-Hilbert action with a cosmological constant $\Lambda$
\begin{equation}
\label{actionWdW:EH}
I_{B} = I_{EH} = \int_{\mathcal{M}}\rmd^D X\sqrt{-g}\,\left(R - 2 \Lambda\right)~.
\end{equation}

The boundaries of the patch are hypersurfaces, which one can define in terms of a scalar function $\Phi(X) =0$. In this paper, we use the convention that $\Phi(X)$ is negative inside the patch and positive outside of it. These conventions agree with those in \cite{Jubb:2016qzt, Carmi:2016wjl} and ensure that the one-form $\rmd \Phi$ always points outward.   
If the hypersurface is space-like or time-like, then the unit normal is taken to be  
\[ 
   n^\alpha = \frac{\partial^\alpha\Phi}{\sqrt{\left|g_{\alpha\beta}
   \partial^\alpha\Phi\partial^\beta\Phi\right|}}~.
\]
In these cases, the boundary contribution to the action is given by the Gibbons-Hawking-York term
\begin{equation}
\label{actionWdW:GHY}
I_{S} = I_{GHY} = 2\int_\mathcal{B}\rmd^{D-1} x\sqrt{|h|}\, K~,
\end{equation}
where $K = \nabla_\alpha n^\alpha$ is the extrinsic curvature of the boundary, and $h$ is the determinant of the induced metric.

Instead, if the boundary is a null hypersurface, then $\partial^\alpha \Phi$ must be proportional to the null tangent vector,
\begin{equation}
\label{actionWdW:vectors}
k^\alpha \equiv \frac{\partial X^\alpha}{\partial\lambda} = e^{\sigma(x)}\partial^\alpha\Phi~.
\end{equation}
Here and henceforth, $\lambda$ denotes the hypersurface coordinate that parameterizes the null direction. The function $\sigma(x)$ determines the parameterization of the null direction, \ie the choice of $\lambda$.  
This has been discussed extensively in \cite{Lehner:2016vdi}. 

The analogue of the Gibbons-Hawking-York term is played by\footnote{The different sign with respect to the expression given in \cite{Lehner:2016vdi} derives from different conventions for $k^\alpha$ and $\Phi$.}   
\begin{equation}
\label{actionWdW:null}
	I_{N} = 2\int \rmd\lambda\rmd^{D-2} x \sqrt{\gamma}\,\kappa~,
\end{equation}
where $\kappa$ is the non-affinity parameter defined by
\begin{equation}
\label{actionWdW:kappa}
k^\beta\nabla_\beta k^\alpha = \kappa k^\alpha~.
\end{equation}
In \eqref{actionWdW:null}, $\gamma$ is the determinant of the induced metric on the codimension-two space-like part of the boundary that is orthogonal to the null direction.

In addition to the codimension-one boundaries, the WdW patch has also codimension-two joints in which two boundaries intersect. The contribution of a joint depends on the type of the intersecting boundaries. In the case of a joint formed by two null boundaries, it is given by 
\begin{equation}
\label{actionWdW:joint}
I_{C} = 2 \epsilon_\mathcal{J} \int_\mathcal{J}\rmd^{D-2} x \sqrt{\gamma}\,\ln\frac{\left|k_1\cdot k_2\right|}{2}~,
\end{equation}
where $k_1$ and $k_2$ are the null tangent vectors of the two intersecting boundaries, and the sign $\epsilon_\mathcal{J} = \pm 1$ depends on the relative position of the boundaries and the bulk region. The (convention-dependent) rules can be found in \cite{Lehner:2016vdi, Carmi:2016wjl}. We remark  that there is no real need to memorize these rules, if one does not fix the parameterization functions $\sigma(x)$. As we shall see below, performing an integration by parts in $I_{N}$, one obtains $\sigma$-dependent contributions localized at the joints, which must cancel against the $\sigma$-dependent parts of the joint term $I_{C}$. This determines the signs $\epsilon_\mathcal{J}$.
We refrain from giving the joint terms in the other cases, because we will not need them.

Adding up all these contributions gives an action with a well-defined action principle. Moreover, the full action is additive, \ie one can safely divide a given space-time region into smaller subregions. However, in the presence of null boundaries, it is not invariant under a reparameterization of the null directions, which can be seen from the fact that the arbitrary functions $\sigma(x)$ remain explicit. A remedy is to add a counterterm such that its variation under a reparameterization will cancel the variation of the action. We will use the term suggested in \cite{Lehner:2016vdi}
\begin{equation}
\label{actionWdW:ct}
I_{\mathrm{c.t.}} =2\int\rmd\lambda\rmd^{D-2} x\sqrt{\gamma}\, \Theta\ln(\tilde{l}\left|\Theta\right|)~,
\end{equation}
where $\Theta$ is the null geodesic expansion. It is defined by $\Theta = \partial_\lambda\ln\sqrt{\gamma}$. The constant $\tilde{l}$ is of dimension length and plays the role of a renormalization scale. 

It has been pointed out in \cite{Jubb:2016qzt} that the choice \eqref{actionWdW:ct} is not unique. There are, in fact, other terms that would serve the same purpose, such as 
\begin{equation}
	\label{actionWdW:ct2}
	I'_{\mathrm{c.t.}} =2\int\rmd\lambda\rmd^{D-2} x\sqrt{\gamma}\, \Theta\ln\frac{\rmd\lambda}{\rmd t}~,
\end{equation}
where $t$ is an arbitrary affine parameterization, or
\begin{equation}
	\label{actionWdW:ct3}
	I''_{\mathrm{c.t.}} =2\int\rmd\lambda\rmd^{D-2} x\sqrt{\gamma}\, \Theta\ln s_{ab}s^{ab}~,
\end{equation}
where $s_{ab}$ is the shear tensor of the null geodesic congruence.
We will use the term \eqref{actionWdW:ct} because, as it turns out, \eqref{actionWdW:ct2} is not compatible with the additive properties of the action, while \eqref{actionWdW:ct3} is not regular on the surfaces we need to study.

\section{Global AdS}
\label{ads}

In this section, we will compute the complexity of pure AdS spacetime, with the intention to find a systematic and well-defined way to set it to zero. The motivation behind our intention is that we want to treat the CFT ground state dual to the AdS vacuum as the reference state, the complexity of which is null by definition. 
Then, we can interpret the holographic complexity of other systems, \eg black holes, obtained using the same prescription, as the complexity relative to the ground state.
Thus, what we are looking for is a way to set the action on the WdW patch in AdS vacuum to zero.   
First, we will see how this can be achieved making use of the parameterization dependence of the null surface terms, if no counter term is included. Then, we show that the counter term \eqref{actionWdW:ct}, which renders the action parameterization independent, is generically divergent. Setting it to zero would require fine tuning the renormalization scale $\tilde{l}$. 

We consider $n+2$-dimensional AdS space-time in global coordinates, with the metric
\begin{equation}
\label{ads:metric}
\rmd s^2 = \frac{L^2}{\cos^2 \rho} \left( -\rmd t^2 + \rmd \rho^2 +\sin^2 \rho \rmd \Omega_n^2 \right)~.
\end{equation}
Here, $L$ is the AdS curvature radius, $\rmd \Omega_n^2$ the metric of a unit $n$-sphere, and the AdS boundary is located at $\rho=\pi/2$. Note that our metric tensor has dimension length$^2$, whereas the coordinates are dimensionless. This requires a slight change with respect to the general construction in the previous section, which we will mention in due course.

The WdW patch is bounded by two null hypersurfaces, which meet at the AdS boundary in a joint. Using the time translation invariance, we are free to choose $t=0$ at this joint. Because of the divergences arising at the boundary, the WdW patch must be regulated. We shall use a regularization in which the radial coordiante of the joint is $\rho_\ast=\pi/2-\varepsilon$. Then, the following two scalar functions define the null boundaries:
\begin{equation}
\label{ads:Phi}	
	\Phi_\pm(t,\rho) = \rho - \rho_\ast \pm t~.
\end{equation}

The contributions to the action to be calculated are the bulk term \eqref{actionWdW:EH}, the non-affinity terms on the null hypersurfaces \eqref{actionWdW:kappa} and the joint term \eqref{actionWdW:joint}. Later, we will also consider the covariance counter term \eqref{actionWdW:ct}. Let us consider them one at a time.
The bulk contribution \eqref{actionWdW:EH} is 
\begin{align}
\notag 
I_{B} &= \int \rmd^{n+2}x \, \sqrt{-g} \left( R-2 \Lambda\right) 
= -\frac{2(n+1)}{L^2} \int \rmd^{n+2}x \, \sqrt{-g} \\
\label{ads:I.B}
&= -4 (n+1) L^n \Omega_n \int\limits_0^{\rho_\ast} \rmd \rho \,\frac{\tan^n\rho}{\cos^2 \rho} 
\left( \rho_\ast - \rho \right)
= - 4 L^n \Omega_n \int\limits_0^{\rho_\ast} \rmd \rho \tan^{n+1} \rho~,
\end{align}
where we have integrated by parts in the last step. $\Omega_n$ denotes the volume of a unit $n$-sphere.

Let us now consider the null surface terms. Following the prescription outlined in Sec.~\ref{actionWdW}, the null tangents are, in coordinates $(t,\rho,\vec{\Omega})$,   
\begin{equation}
\label{ads:k}
k_\pm^\mu = \e{\sigma_\pm} \frac{\cos^2\rho}{L}  \left( \mp 1, 1, \vec{0} \right)~, 
\end{equation}
where $\sigma_\pm$ are two auxiliary functions of $\lambda$ incorporating the freedom of parameterization of the null directions. We remark that we have  multiplied the expression resulting from \eqref{actionWdW:vectors} by a factor of $L$, because our space-time coordinates, according to \eqref{ads:metric}, are dimensionless, while the coordinates in \eqref{actionWdW:vectors} have dimension of length.  
Notice that \eqref{ads:k} implies 
\begin{equation}
\label{ads:drho.dl}
   \frac{\partial\rho}{\partial\lambda} = \frac{\cos^2\rho}{L} \e{\sigma_\pm}~.
\end{equation}
Because this is positive, $\lambda$ increases towards the corner on both boundaries. 

The non-affinity parameters $\kappa_\pm$ turn out to be $\kappa_\pm = \partial_{\lambda} \sigma_\pm$, so that the action term $I_N$ is the sum of
\begin{align}
\notag
I_{N\pm} &= 2 L^n \Omega_n \int \rmd\lambda\, \tan^n \rho \, \partial_\lambda \sigma_\pm \\
\label{ads:I.N}
&= 2 L^n \Omega_n \left[ \sigma_\pm(\rho_\ast) \tan^n\rho_\ast- 
\int\limits_0^{\rho_\ast} \rmd\rho\, \sigma_\pm \frac{ n \tan^{n-1}\rho}{\cos^2 \rho} \right]~.
\end{align} 
We have integrated by parts, and $\sigma_\pm(\rho_\ast)$ denotes the values of $\sigma_\pm$ at the joint.  

The corner contribution is simply
\begin{equation}
\label{ads:I.C}
I_{C} = -2  L^n \Omega_n \tan^n \rho_\ast \ln \frac{k_+\cdot k_-}{2} = -2 L^n \Omega_n \tan^n \rho_\ast
\left[ \sigma_+(\rho_\ast) + \sigma_-(\rho_\ast) + 2 \ln \cos \rho_\ast \right]~. 
\end{equation}
We remark that the sign of the corner term follows easily from the fact that the terms with $\sigma_\pm(\rho_\ast)$ must cancel between $I_{N\pm}$ and $I_{C}$. 

Adding the contributions \eqref{ads:I.B}, \eqref{ads:I.N} and \eqref{ads:I.C} yields the action without the covariance term,
\begin{equation}
\label{ads:I.WdW.no.cov}
I_{B} + I_{N} + I_{C} = - 4 L^n \Omega_n  \int\limits_0^{\rho_\ast} \rmd \rho \left[ \tan^{n+1}\rho + \frac{n}{2}(\sigma_+ + \sigma_-)
\frac{\tan^{n-1}\rho}{\cos^2 \rho}  \right] 
- 4 L^n \Omega_n \tan^n \rho_\ast \ln \cos\rho_\ast~.
\end{equation}
We can further rewrite the integral from the bulk contribution as 
\begin{align}
\notag 
\int\limits_0^{\rho_\ast} \rmd \rho\, \tan^{n+1}\rho &= - \int\limits_0^{\rho_\ast} \rmd \rho\, \tan^n \rho\, 
\partial_\rho \ln \cos\rho  \\
\label{ads:massage}
&= - \tan^n \rho_\ast \ln \cos \rho_\ast + n \int\limits_0^{\rho_\ast} \rmd \rho\, 
\frac{\tan^{n-1}\rho}{\cos^2 \rho} \ln \cos \rho~,
\end{align}
so that \eqref{ads:I.WdW.no.cov} becomes
\begin{equation}
\label{ads:I.WdW.no.cov.final}
I_{B} + I_{N} + I_{C} = - 2 n L^n \Omega_n \int\limits_0^{\rho_\ast} \rmd \rho\, \frac{\tan^{n-1}\rho}{\cos^2 \rho} 
\left( \sigma_+ + \sigma_- + 2\ln \cos\rho \right)~.
\end{equation}
We note that, written in this form, the integrand that gives the total action is proportional to a term that has the same structure as the corner term. We also remark that the dependencies of $\sigma_+$ and $\sigma_-$ on $\rho$ are to be intended implicitly along the null boundaries to the future and the past of the WdW patch, respectively.

It is now obvious that there exists a class of parameterizations, for which \eqref{ads:I.WdW.no.cov.final} vanishes. For example, one may choose  
\begin{equation}
\label{ads:choice}
  \text{on $N+$:} \quad \sigma_+(\lambda) = a_+ \ln \cos \rho(\lambda)~, \qquad
  \text{on $N-$:} \quad \sigma_-(\lambda) = a_- \ln \cos \rho(\lambda)~,
\end{equation}
where $a_+$ and $a_-$ are two constants satisfying the constraint $a_+ + a_- = 2$. We also see that the corner term \eqref{ads:I.C} vanishes separately with such a choice.

As we mentioned at the beginning of this section, a vanishing action on the WdW patch is exactly what we were looking for in the case of pure AdS. Nevertheless, let us also discuss the covariance term \eqref{actionWdW:ct}, which, for general parameterization, is
\begin{equation}
\label{ads:I.WdW.aux}
	I_{\mathrm{c.t.}} = 2 L^n \Omega_n  \int\limits_0^{\rho_\ast} \rmd \rho\, ( \partial_\rho \tan^n \rho )
\left( 2 C - 2\ln \tan \rho + \sigma_+ + \sigma_- \right)~.
\end{equation}
Here, we have abbreviated 
\begin{equation}
\label{ads:C.def}
C = \ln \frac{n\tilde{l}}{L}~.
\end{equation}
For general parameterizations, the terms with $\sigma_\pm$ in \eqref{ads:I.WdW.aux} cancel those in \eqref{ads:I.WdW.no.cov.final} as expected. 

If we adopt the parameterization \eqref{ads:choice}, for which the counter term  represents the entire parameterization-invariant action, \eqref{ads:I.WdW.aux} becomes  
\begin{align}
\label{ads:I.WdW.final} 
I_{\mathrm{c.t.}} &= 
4 L^n \Omega_n \left[ \tan^n \rho_\ast \left( C - \ln \sin \rho_\ast \right) + 
\int\limits_0^{\rho_\ast} \rmd \rho \tan^{n-1} \rho \right]~.
\end{align}
The integral on the right hand side of \eqref{ads:I.WdW.final} can be evaluated in closed form, 
\begin{equation}
\label{ads:integral}
\int\limits_0^{\rho_\ast} \rmd \rho \tan^{n-1} \rho = \sum\limits_{k=1}^{\left[\frac{n-1}2\right]} (-1)^{k-1} 
\frac{\tan^{n-2k} \rho_\ast}{n-2k} + (-1)^{\left[\frac{n}2\right]} 
\begin{cases}
\ln \cos \rho_\ast \quad &\text{for even $n$,}\\
\rho_\ast &\text{for odd $n$.}
\end{cases}
\end{equation}

Clearly, \eqref{ads:I.WdW.final} is generically divergent for $\rho_\ast \to \pi/2$. In order to get a vanishing result, one could interpret the renormalization scale $\tilde{l}$ or, equivalently, the renormalization constant $C$, as a function of the cut-off $\rho_\ast$ and fine tune it such that \eqref{ads:I.WdW.final} vanishes. The choice $C=0$, which was adopted in \cite{Akhavan:2019zax}, only removes the leading divergence.

\section{AdS-Schwarzschild black hole}
\label{sads}

\subsection{Setup}

In this section, we will consider the case of AdS-Schwarzschild space-time with a spherical horizon. This case was discussed in \cite{Brown:2015bva, Brown:2015lvg, Lehner:2016vdi, Carmi:2017jqz} and more recently in \cite{Akhavan:2019zax}. However, in most of these papers, only the time evolution of complexity, $\partial \mathcal{C}/\partial t$, was discussed, because only the difference of the actions in two slightly different WdW patches is needed in such a calculation and the divergent constant drops out.
Here, we will consider the entire WdW patch and compute the complexity of the black hole with respect to the pure AdS geometry.\\

AdS-Schwarzschild space-time is described by the metric
\begin{equation}
\label{sads:metric}
\rmd s^2 = -f(r) \rmd t^2 +f(r)^{-1} \rmd r^2 + r^2 \rmd \Omega_n^2~,
\end{equation}
with 
\begin{equation}
\label{sads:f}
f(r) = \frac{r^2}{L^2} + 1 -\frac{\omega^{n-1}}{r^{n-1}}~.
\end{equation}
The parameter $\omega$ is related to the total mass 
\begin{equation}
\label{sads:M}
	M = \frac{n\Omega_n}{16\pi G} \omega^{n-1}~.
\end{equation} 
The horizon radius is defined by $f(r_H)=0$. The black hole temperature is
\begin{equation}
\label{sads:temp}
	T_{BH}=\frac{1}{4\pi}\left(\frac{n+1}{L^2}r_H+\frac{n-1}{r_H}\right)~,
\end{equation}
and the entropy is given by the Bekenstein-Hawking formula
\begin{equation}
\label{sads:entropy}
S=\frac{A}{4\hbar G}~,
\end{equation}
where $A=\Omega_n r_H^n$ is the spatial area of the black hole horizon.

The AdS-Schwarzschild black hole has also a time scale relevant to the time evolution of the complexity know as the scrambling time \cite{Hayden:2007cs, Sekino:2008he}. The scrambling time of a system is a measure of how fast said system can thermalize information by means of the interactions between its elementary degrees of freedom. For the case of a black hole, the scrambling time is the time it takes the black hole to completely and uniformly spread a local perturbation across its horizon. Interestingly enough, black holes are believed to be the fastest scramblers in nature and their scrambling time is of the order
\begin{equation}
\label{sads:scram}
t_{\ast}\sim\beta_{BH}\ln S~.
\end{equation}

\begin{figure}[htp]
	\begin{center}
		\includegraphics[width=.45\textwidth,align=c]{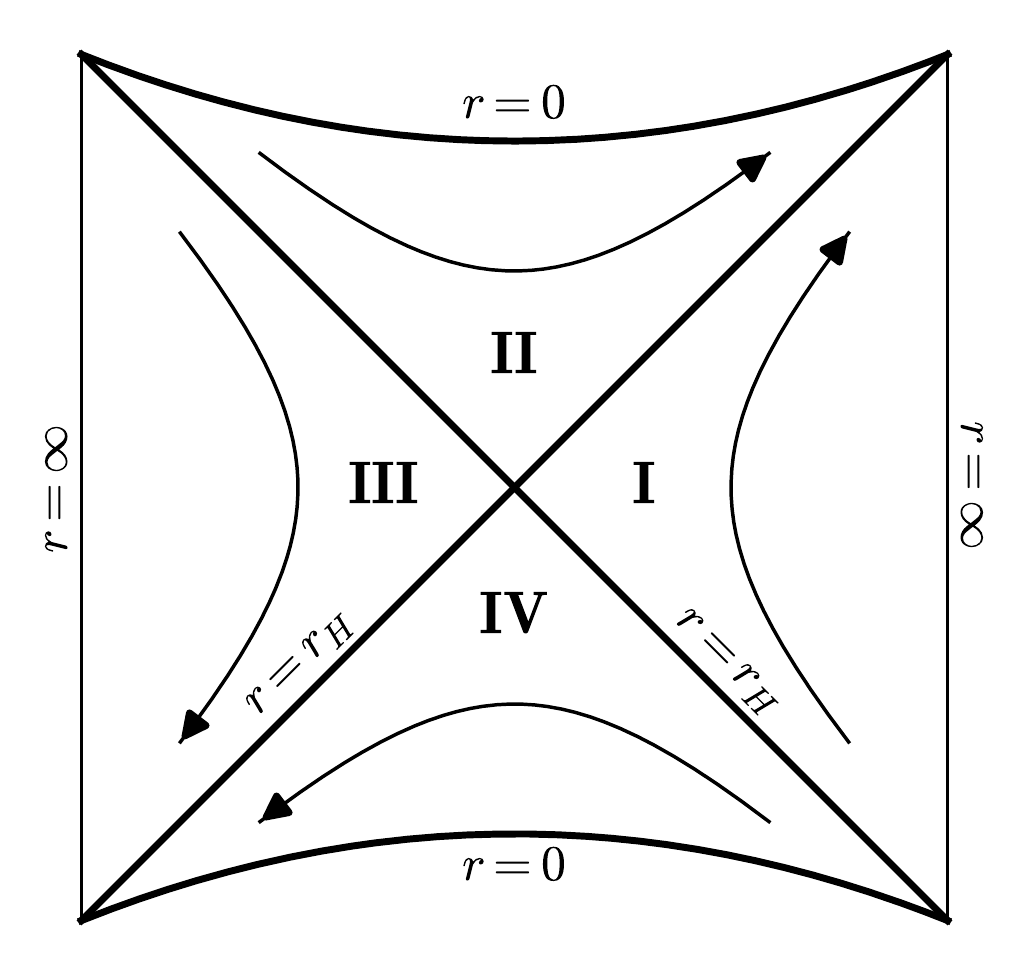}%
		\hfill%
		\caption{Penrose diagram of the maximally extended eternal AdS-Schwarzschild space-time. The arrows indicate the flow of the Killing vector field $\partial_t$. \label{fig:schwads}}
	\end{center}
\end{figure}

The Penrose diagram of maximally extended AdS-Schwarzschild space-time is shown in Fig.~\ref{fig:schwads}. Each of the quadrants $\mathrm{I}$--$\mathrm{IV}$ is covered by a set of coordinates $(t,r)$ with metric \eqref{sads:metric}, where $r>r_H$ in quadrants $\mathrm{I}$ and $\mathrm{III}$ and $r<r_H$ in $\mathrm{II}$ and $\mathrm{IV}$ and the future and past singularities are situated at $r=0$. 
For the upcoming computations it will be useful to work with Eddington-Finkelstein coordinates. We define the tortoise coordinate by
\begin{equation}
\label{sads:tortoise}
	r^\ast(r) = \int\limits_{R}^r \frac{\rmd r}{f(r)}~.
\end{equation} 
Note our choice of the lower integration limit, where $R$ will be taken to agree with the cut-off instead of the conventional $\infty$. This will somewhat simplify the calculations.  

With ingoing Eddington-Finkelstein coordinates, the metric is 
\begin{equation}
\label{sads:metricVR}
\rmd s^2 = -f(r)\rmd v^2 + 2\rmd v\rmd r + r^2\rmd\Omega_n^2~,
\end{equation}
where $v = t+ r^\ast$. The ingoing Eddington-Finkelstein coordinates cover two quadrants, $\mathrm{I} \cup\mathrm{II}$ or $\mathrm{III} \cup \mathrm{IV}$.
Likewise, with outgoing coordinates, the metric is 
\begin{equation}
\label{sads:metricUR}
\rmd s^2 = -f(r)\rmd u^2 - 2\rmd u\rmd r + r^2\rmd\Omega_n^2~,
\end{equation}
where $u = t -r^\ast$. The outgoing Eddington-Finkelstein coordinates cover two quadrants, $\mathrm{I} \cup\mathrm{IV}$ or $\mathrm{II} \cup \mathrm{III}$.

\begin{figure}[th]
	\begin{center}
		\includegraphics[width=.45\textwidth,align=c]{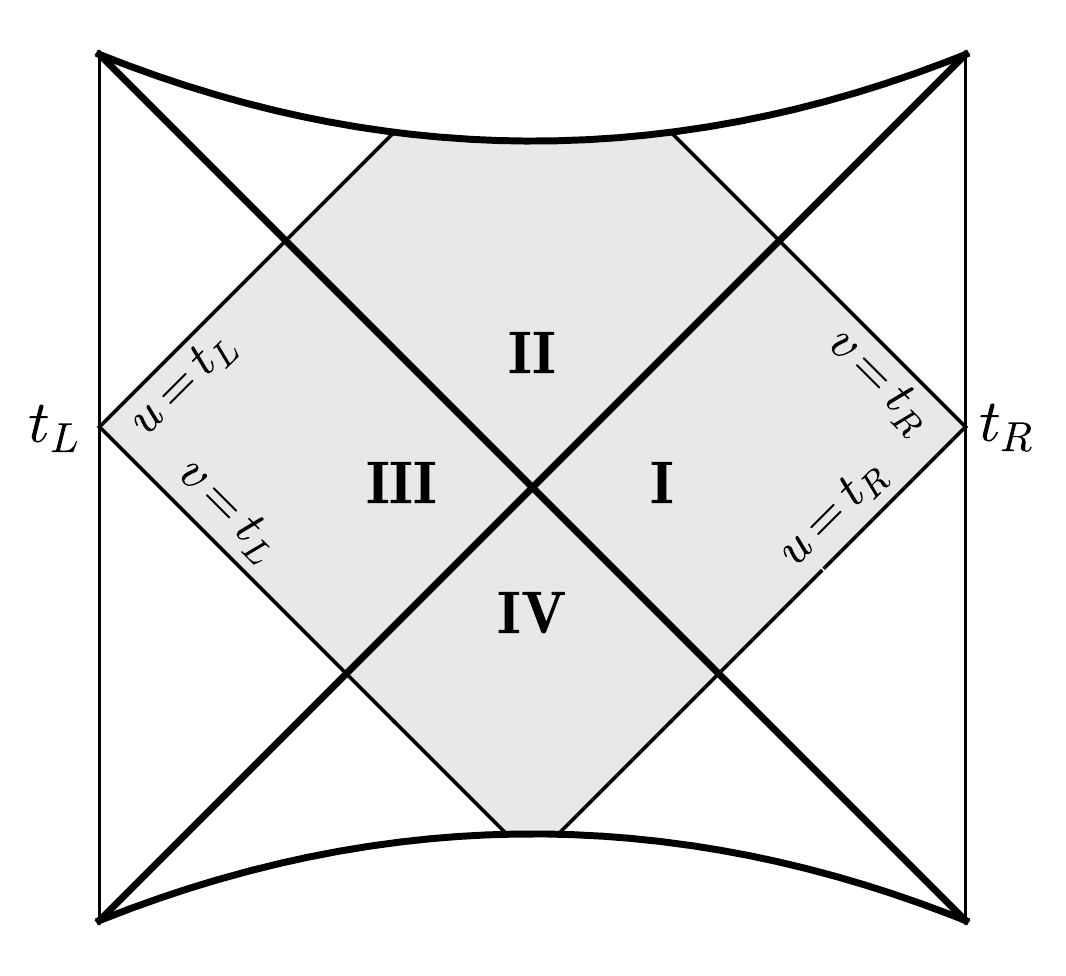}%
		\hfill%
		\includegraphics[width=.45\textwidth,align=c]{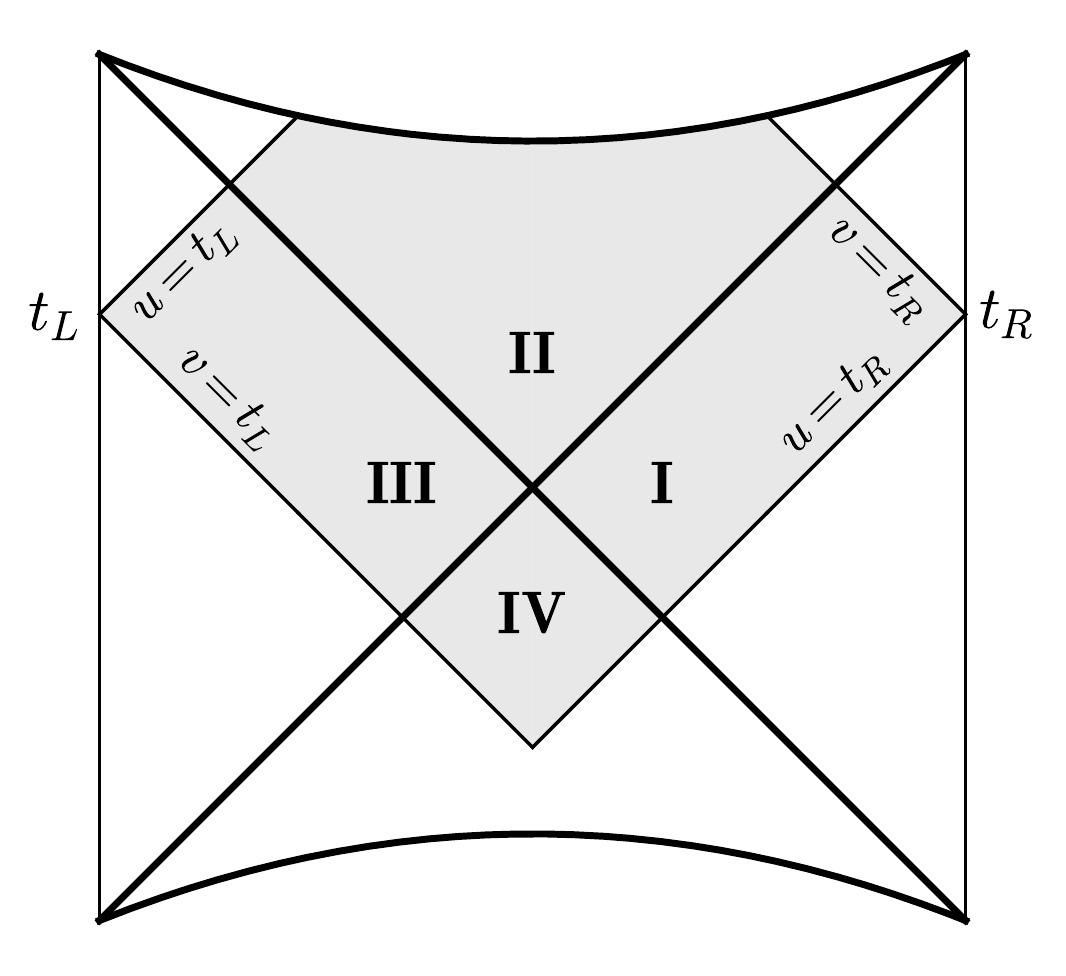}
		\caption{Left: WdW patch for the case $-\tau_0<\tau<\tau_0$. The WdW patch touches both future and past singularities.  
			Right: WdW patch for $\tau>\tau_0$. Two null boundaries meet in the quadrant IV. \label{fig:wdw}}
	\end{center}
\end{figure}

The WdW patch is bounded by the null surfaces intersecting the left and right boundaries at the cut-off radius $R$ and at times $t_L$ and $t_R$, respectively, and possibly by parts of the future and past singularities. The cut-off radius $R$ will be taken to infinity at the end. 
Furthermore, to simplify, we use time translational invariance to set $t_R=-t_L=\tau$.
The precise shape of the WdW patch depends on the value of $\tau$. When $|\tau|<\tau_0= -r^\ast(0)$, the WdW patch touches both, the future and past singularities. For $\tau > \tau_0$, the WdW patch touches only the future singularity, and two null boundaries meet in the quadrant $\mathrm{IV}$. Let us call $r_m$ the radius at which they meet. It is given implicitly by the relation $r^\ast(r_m)=-\tau$. Similarly, for $\tau<-\tau_0$, two boundaries meet in the quadrant II and the WdW patch touches only the past singularity. Because the setup is symmetric under $\tau\to -\tau$, this last situation does not need to be discussed separately.
The WdW patch in the other two cases are illustrated in Fig.~\ref{fig:wdw}.

\subsection{Action in a causal diamond}
\label{sads:caus.diam}

\begin{figure}[th]
	\begin{center}
		\includegraphics[width=.45\textwidth,align=c]{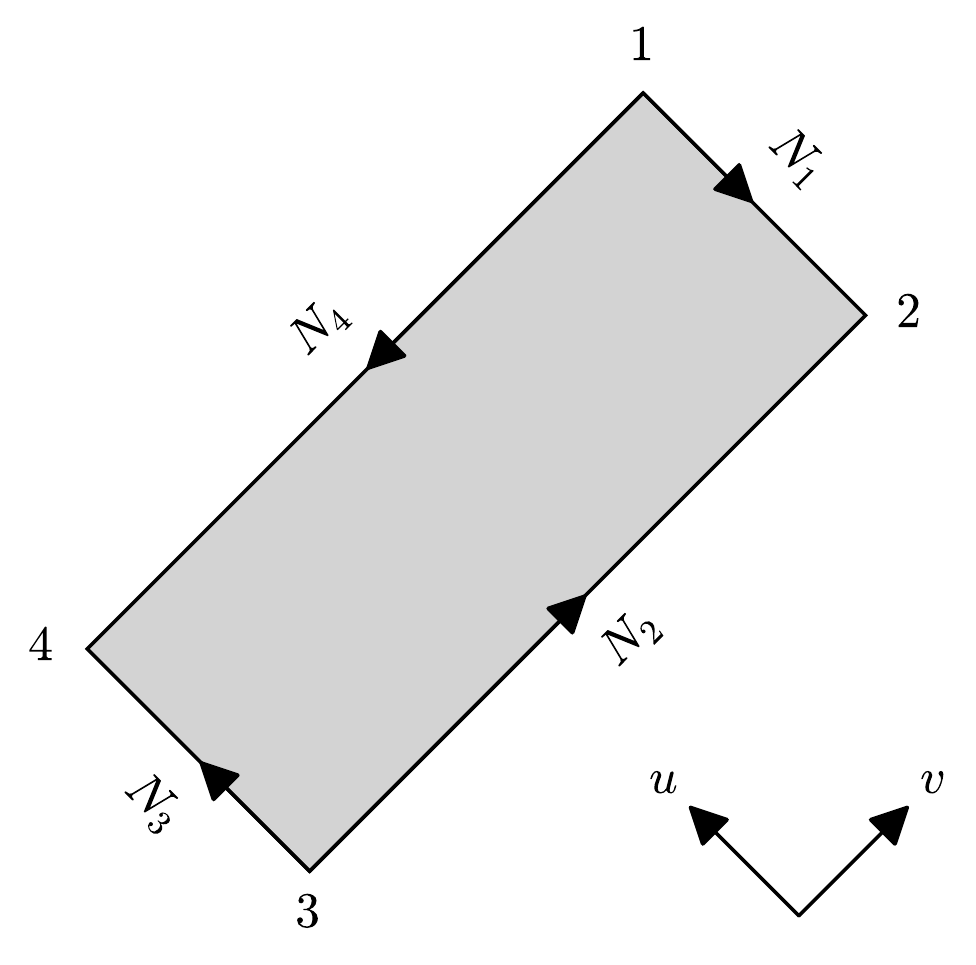}%
		\caption{A generic causal diamond.
		The labels of the null boundaries and the corners used in the text are shown. 
		The arrows on the null boundaries indicate the orientation of the $\lambda$-integrals in 
		the corresponding boundary terms of the action.
		\label{fig:caus.diamond}}
	\end{center}
\end{figure}

We are interested in the action in the WdW patch, which represents the complexity of the dual state. Before considering the entire WdW patch, let us focus on a causal diamond, which we may place, without loss of generality, in the quadrant $\mathrm{I}$. The causal diamond is bounded by four null surfaces, which we label $N_1$, \ldots, $N_4$, counting them clockwise starting from the north east. The four intersection points are counted clockwise starting from the north.  Obviously, their coordinates satisfy $v_1 = v_2$, $u_2=u_3$, $v_3=v_4$ and $u_4 = u_1$. 
The setup is illustrated in Fig.~\ref{fig:caus.diamond}.
In what follows, we work in outgoing Eddington-Finkelstein coordinates \eqref{sads:metricUR}. 

The four scalar functions defining the null surfaces are given by
\begin{subequations}
\begin{align}
	\label{sads:phi1}
		\Phi_1(u,r) &= u+ 2r^\ast(r) - v_1~,\\
	\label{sads:phi2}
		\Phi_2(u,r) &= u_2 -u~,\\
	\label{sads:phi3}
		\Phi_3(u,r) &= v_3 - u- 2r^\ast(r)~,\\
	\label{sads:phi4}
		\Phi_4(u,r) &= u - u_4~.
\end{align}
\end{subequations}
From these, we obtain the following expressions for the null tangent vectors,
\begin{subequations}
\begin{align}
	\label{sads:k1}
		k_1^\alpha &= \e{\sigma_1} \left( -\frac{2}{f}, 1 , \vec{0} \right)~,\\
	\label{sads:k2}
		k_2^\alpha &= \e{\sigma_2} \left( 0, 1 , \vec{0} \right)~,\\
	\label{sads:k3}
		k_3^\alpha &= \e{\sigma_3} \left( \frac{2}{f}, -1 , \vec{0} \right)~,\\
	\label{sads:k4}
		k_4^\alpha &= \e{\sigma_4} \left( 0, -1 , \vec{0} \right)~.
\end{align}
\end{subequations}
The four functions $\sigma_1$, \ldots, $\sigma_4$ implement the parameterization dependence. 

Let us start with the surface terms \eqref{actionWdW:null}. For all four surfaces, we have $\kappa = \frac{\rmd \sigma}{\rmd \lambda}$. The orientation of the $\lambda$-integrals can be read off from \eqref{sads:k1}--\eqref{sads:k4}, because $k^\alpha = \frac{\rmd x^\alpha}{\rmd \lambda}$. For example, for $N_1$ we get
\begin{equation}
\label{sads:IN1}
	\frac{I_{N_1}}{2\Omega_n} = \int\rmd\lambda\, r^n\kappa 
	= \int\limits_{r_1}^{r_2}\rmd r\,r^n\frac{\rmd\sigma_1}{\rmd r} 
	= -n\int\limits_{r_1}^{r_2}\rmd r\,r^{n-1}\sigma_1 + r_2^n\sigma_1(r_2) - r_1^n\sigma_1(r_1)~.
\end{equation}
Proceeding similarly for the other three boundaries and summing up all the terms, we find
\begin{align}
\label{sads:IN}
	\frac{I_{N}}{2\Omega_n} &= 
	-n\int\limits_{r_1}^{r_2}\rmd r\,r^{n-1}\sigma_1
	-n\int\limits_{r_3}^{r_2}\rmd r\,r^{n-1}\sigma_2 
	-n\int\limits_{r_3}^{r_4}\rmd r\,r^{n-1}\sigma_3 
	-n\int\limits_{r_1}^{r_4}\rmd r\,r^{n-1}\sigma_4 \\
\notag
	&\quad 
	- r_1^n\left[\sigma_1(r_1)+\sigma_4(r_1)\right] + r_2^n\left[\sigma_1(r_2)+\sigma_2(r_2)\right] 
	- r_3^n\left[\sigma_2(r_3)+\sigma_3(r_3)\right] + r_4^n\left[\sigma_3(r_4)+\sigma_4(r_4)\right]~.
\end{align}

The bulk action \eqref{actionWdW:EH} gives
\begin{equation}
\label{sads:IB-1} 
	\frac{I_{B}}{2\Omega_n} = - \frac1{L^2} \int\limits_{u_2}^{u_4} \rmd u 
		\left[ \rho_1(u)^{n+1} - \rho_3(u)^{n+1} \right]~,
\end{equation}
where the functions $\rho_1(u)$ and $\rho_3(u)$ are defined implicitly by $\Phi_1(u,\rho_1)=0$ and $\Phi_3(u,\rho_3)=0$, respectively. Using 
\[ 
	\frac{\rho}{L^2} = \frac12 
	\left[ \frac{\rmd f(\rho)}{\rmd \rho} - (n-1) \frac{\omega^{n-1}}{\rho^n} \right]
	= -\frac{\rmd}{\rmd u} \ln |f(\rho)| - \frac12(n-1)\frac{\omega^{n-1}}{\rho^n}~,
\]
we can rewrite \eqref{sads:IB-1} as
\begin{equation}
\label{sads:IB-2} 
	\frac{I_{B}}{2\Omega_n} = \int\limits_{u_2}^{u_4} \rmd u 
		\left[ \rho_1^n \frac{\rmd}{\rmd u} \ln |f(\rho_1)| 
		- \rho_3^n \frac{\rmd}{\rmd u} \ln |f(\rho_3)| \right]~.
\end{equation}
After integrating by parts and changing the integration variable, this becomes
\begin{align}
\label{sads:IB-3} 
	\frac{I_{B}}{2\Omega_n} 
		&= n \int\limits_{r_1}^{r_2} \rmd r\, r^{n-1} \ln |f(r)| 
		+ n \int\limits_{r_3}^{r_4} \rmd r\, r^{n-1} \ln |f(r)| \\
\notag &\quad
		+ r_1^n \ln |f(r_1)| - r_2^n \ln |f(r_2)|
		+ r_3^n \ln |f(r_3)| - r_4^n \ln |f(r_4)|~.
\end{align}
This is identical to 
\begin{align}
\label{sads:IB}
	\frac{I_B}{2\Omega_n} 
		&= n a_u\int\limits_{r_1}^{r_2}\rmd r\,r^{n-1}\ln|f(r)| 
		 + n a_u\int\limits_{r_3}^{r_4}\rmd r\,r^{n-1}\ln|f(r)| \\
	\notag &\quad
		+ n a_v\int\limits_{r_3}^{r_2}\rmd r\,r^{n-1}\ln|f(r)| 
		+ n a_v\int\limits_{r_1}^{r_4}\rmd r\,r^{n-1}\ln|f(r)| \\
	\notag &\quad 
		+ r_1^n\ln|f(r_1)|-r_2^n\ln|f(r_2)|+r_3^n\ln|f(r_3)|-r_4^n\ln|f(r_4)|~,
\end{align}
where $a_u$ and $a_v$ are two real constants that are constrained by $a_u + a_v = 1$.

The corner terms \eqref{actionWdW:joint} contribute
\begin{align}
\label{sads:IC}
	\frac{I_C}{2\Omega_n} 
		&= r_1^n \left[\sigma_1(r_1)+\sigma_4(r_1)-\ln|f(r_1)|\right] 
		- r_2^n \left[\sigma_1(r_2)+\sigma_2(r_2)-\ln|f(r_2)|\right] \\ 
\notag &\quad 
		+ r_3^n \left[\sigma_2(r_3)+\sigma_3(r_3)-\ln|f(r_3)|\right] 
		- r_4^n \left[\sigma_3(r_4)+\sigma_4(r_4)-\ln|f(r_4)|\right]~.
\end{align}
It is apparent that we have manipulated the surface and bulk actions in such a way that the boundary terms arising from the integrations by parts precisely cancel the corner contribution. Thus, after summing \eqref{sads:IN}, \eqref{sads:IB} and \eqref{sads:IC}, the action of a causal diamond is obtained as 
\begin{align}
\label{sads:CD}
	\frac{I_B+I_N+I_C}{2\Omega_n} 
	&= n\int\limits_{r_1}^{r_2}\rmd r\,r^{n-1}\left[ a_u\ln|f(r)|-\sigma_1\right] 
	+ n\int\limits_{r_3}^{r_2}\rmd r\,r^{n-1}\left[a_v\ln|f(r)|-\sigma_2\right] \\
\notag &\quad 
	+ n\int\limits_{r_3}^{r_4}\rmd r\,r^{n-1}\left[a_u\ln|f(r)|-\sigma_3\right] 
	+ n\int\limits_{r_1}^{r_4}\rmd r\,r^{n-1}\left[a_v\ln|f(r)|-\sigma_4\right]~.
\end{align}
Here, we note that the integral involving $\sigma_1$ must proceed along the null surface $N_1$, because $\sigma_1$ is defined only on $N_1$, and similarly for the others. 

Now, we can require that the action in any causal diamond (of a vacuum region) vanishes. This means that we must choose a parameterization such that
\begin{equation}
\label{sads:choice}
\begin{aligned}
	\text{on $N_1$:}\quad \sigma_1(\lambda) &= a_u \ln|f(r(\lambda))|~, \qquad 
	\text{on $N_3$:}\quad \sigma_3(\lambda)  = a_u \ln|f(r(\lambda))|~,\\
	\text{on $N_2$:}\quad \sigma_2(\lambda) &= a_v \ln|f(r(\lambda))|~, \qquad 
	\text{on $N_4$:}\quad \sigma_4(\lambda)  = a_v \ln|f(r(\lambda))|~.
\end{aligned}
\end{equation}
Interestingly, because $a_u+a_v=1$, this choice implies that all corner terms in \eqref{sads:IC} vanish separately. 

\begin{figure}[th]
	\begin{center}
		\includegraphics[width=.45\textwidth,align=c]{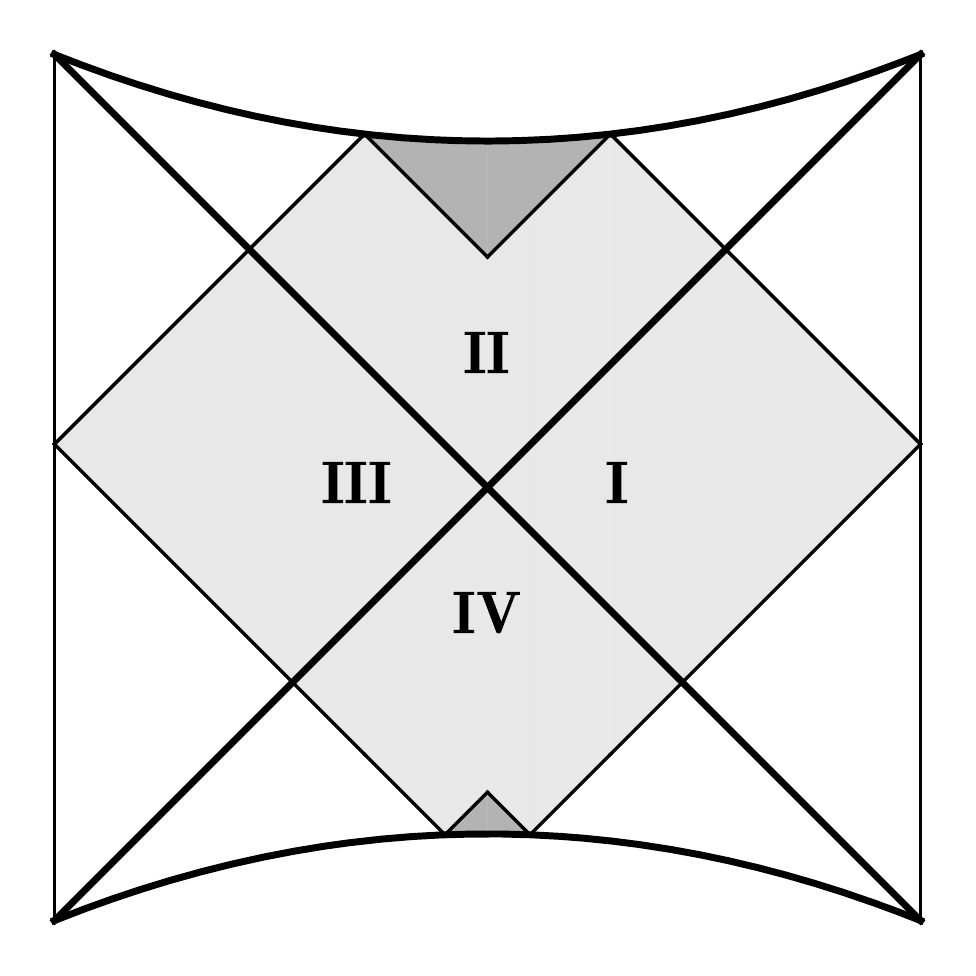}%
		\hfill%
		\includegraphics[width=.45\textwidth,align=c]{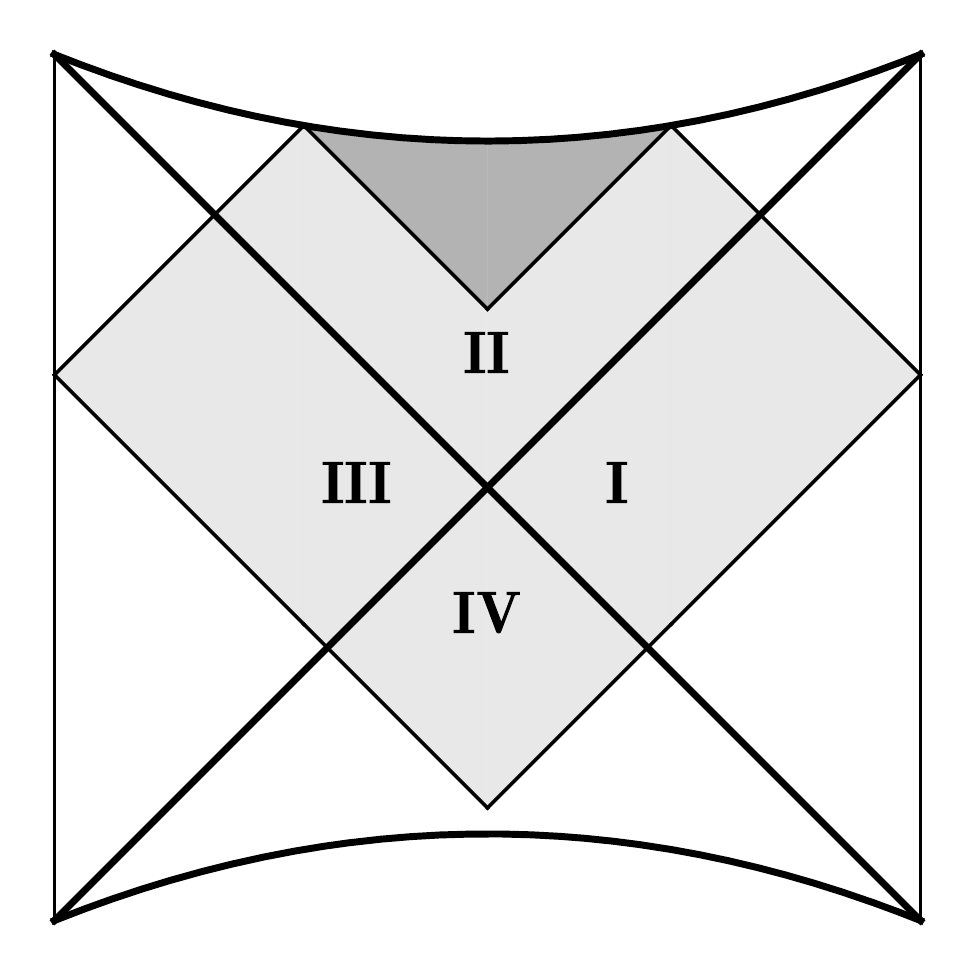}
		\caption{The only portions of the WdW patch that contribute to the action in our computation are those 	shaded in dark gray.\label{fig:wdw_dark}}
	\end{center}
\end{figure}

Our choice \eqref{sads:choice} has an important consequence for the complexity calculation. Using the additivity of the action,  we now know that portions of the WdW patch that are bounded only by null segments do not contribute to the action. Non-zero contributions may come only from the remaining regions, which border with the singularity. These are the dark shaded areas in Fig.~\ref{fig:wdw_dark}. 
Notably, these areas lie entirely behind the horizon. 
We will now turn to the action on these remaining regions. 

\subsection{Action on the WdW patch}
\label{sads:action.wdw}

We have seen in the previous subsection that the calculation of the action can be reduced to the regions that border with the past and future singularities. We first recall that the time translation symmetry is fixed by setting $t_R=-t_L=\tau$. Let us start with the case $\tau>\tau_0=-r^\ast(0)$. In this case, there is only one contributing region, because the WdW patch touches only the future singularity. It is bounded by the two null surfaces $u=t_R+2\tau_0$ and $v=t_L - 2\tau_0$, as well as the part of the future singularity with $t_L-\tau_0 \leq t \leq t_R + \tau_0$. In outgoing Eddington-Finkelstein coordinates, the bulk region is given by $t_L \leq u \leq t_R + 2 \tau_0$ and $0<r\leq \rho(u)$, where $\rho(u)$ is determined implicitly by 
\begin{equation}
\label{sads:rho.def}
	u + 2 r^\ast(\rho) = t_L -2 \tau_0~. 
\end{equation} 

The bulk term is easily computed,
\begin{equation}
\label{sads:IBC}
	\frac{I_B}{2\Omega_n} = -\frac1{L^2} \int\limits_{t_L}^{t_R+2 \tau_0} \rmd u\, [\rho(u)]^{n+1}~.
\end{equation}
Now consider the null surface contribution. Without loss of generality, let us pick the values $a_u=1$, $a_v=0$ in \eqref{sads:choice}. Then, only the null surface with constant $v$ contributes, with $\sigma= \ln (-f)$. 
Its tangent vector is 
\begin{equation}
\label{sads:k5}
		k^\alpha = \e{\sigma} \left( \frac{2}{f}, -1, \vec{0} \right) = \left(-2,f,\vec{0}\right)~.
\end{equation} 
This gives 
\begin{align}
\notag
	\frac{I_N}{2\Omega_n} &= \int \rmd \lambda\, r^n \frac{\rmd}{\rmd \lambda} \ln(-f) 
	= -\int\limits_{t_L}^{t_R+2 \tau_0} \rmd u\, [\rho(u)]^n \frac{\rmd}{\rmd u} \ln(-f) 
	= \frac12 \int\limits_{t_L}^{t_R+2 \tau_0} \rmd u\, \rho^n \frac{\rmd}{\rmd \rho} f(\rho) \\
	&=  \frac1{L^2} \int\limits_{t_L}^{t_R+2 \tau_0} \rmd u\, [\rho(u)]^{n+1} 
		+ \frac12 (t_R-t_L+2\tau_0) (n-1)\omega^{n-1}~.
	\label{sads:INC}
\end{align}

As we mentioned at the end of the previous subsection, the corner term from the joint between the two null surfaces vanishes in our choice of parameterization. The two corners at the future singularity do not contribute either, because their volumes are zero. The remaining contribution is the Gibbons-Hawking-York term \eqref{actionWdW:GHY} at the singularity. Consider a surface with small, but constant $r$, which will be sent to zero at the end. In Schwarzschild coordinates, the normalized, outward-pointing normal vector on this surface is $n^\alpha = ( 0, \sqrt{-f},\vec{0})$, so that the extrinsic curvature reads
\begin{equation}
\label{sads:K}
	K = \frac1{r^n} \partial_r \left(r^n \sqrt{-f}\right) 
	= \frac{n}r \sqrt{-f} - \frac1{\sqrt{-f}} \left[ \frac{r}{L^2} +\frac{(n-1)\omega^{n-1}}{2r^n} \right]~.
\end{equation}
Therefore, the Gibbons-Hawking-York term is 
\begin{equation}
\label{sads:IS}
	\frac{I_S}{2\Omega_n} = \int\limits_{t_L-\tau_0}^{t_R+\tau_0} \rmd t\, 
	\lim_{r\to 0} \left( r^n \sqrt{-f} K \right)
	= \frac12  (t_R-t_L+2\tau_0) (n+1)\omega^{n-1}~.
\end{equation}   

Adding up all contributions yields 
\begin{equation}
\label{sads:I.WdW}
	I_{\text{WdW}} = I_B+I_N+I_C+I_S = 4 n \Omega_n \omega^{n-1} (\tau+\tau_0)~,  
\end{equation}
where we have set $t_R=-t_L = \tau$ in the final result. 

Eqn.\ \eqref{sads:I.WdW} holds for $\tau>\tau_0$ only. The other cases can be obtained using the $t \to -t $ symmetry. In particular, for $|\tau|<\tau_0$, we need to add the contribution from the region bordering with the past singularity. Its contribuition is simply \eqref{sads:I.WdW} with $\tau$ replaced by $-\tau$. The sum of the  contributions from the two regions results in a constant, $I_{\text{WdW},0} = 8 n \Omega_n \omega ^{n-1} \tau_0$.

We are now in a position to translate our result into a complexity. Using \eqref{intro:complexity} and $I =16\pi G S$, we obtain
\begin{equation}
\label{sads:complexity}
	\mathcal{C} =  \frac{4M}{\pi \hbar} 
		\begin{cases}
			|\tau| + \tau_0 \quad &\text{for $|\tau| >\tau_0$,}\\
			2 \tau_0 &\text{for $|\tau| \leq \tau_0$.}
		\end{cases}
\end{equation}
To conclude this subsection, let us discuss what we have found. First, the complexity is manifestly finite. This is ensured by the fact that the corners at the cut-off boundary, where divergences may occur, belong to causal diamonds, which do not contribute following our prescription. No further counter term is needed. We will see in the next subsection that including the covariance counter term re-introduces the generic divergence. The limit $R\to \infty$ has not been taken explicitly. In fact, the only $R$-dependence is hidden in $\tau_0=-r^\ast(0)$, which includes $R$ as the lower integration limit in \eqref{sads:tortoise}. $\tau_0$ remains finite in the limit $R\to\infty$. For example, for $n = 2$ we have
\begin{align}
	\notag
	\tau_0 &= \frac{L^2}{3r_H^2+L^2}\left[\frac{r_H}{2}\ln\left(1+\frac{L^2}{r_H^2}\right)+\frac{3r_H^2+2L^2}{\sqrt{3r_H^2+4L^2}}\arctan\frac{\sqrt{3r_H^2+4L^2}}{r_H}\right]\\
	\label{sads:tau0}
	&=L\frac{\alpha}{1+3\alpha^{2}}\left[\frac{1}{2}\ln\left(1+\alpha^{-2}\right)+\frac{2+3\alpha^2}{\sqrt{4+3\alpha^2}}\arctan\sqrt{3+4\alpha^{-2}}\right]~.
\end{align}
On the second line we have introduced the adimensional ratio $\alpha = \frac{r_H}{L}$. We will always assume $\alpha>1$ (when $\alpha<1$ the black hole is not stable).

Our second observation is that the complexity grows linearly for $\tau>\tau_0$ and saturates the Lloyd bound \cite{Lloyd:2000, Brown:2015lvg, Brown:2015bva}\footnote{To compare with the standard expression for Lloyd's bound, one should set $t=2\tau$.}. Compare this to the reparameterization-invariant approach, which we will review in the next subsection. There, linear growth holds only at late times.
Third, \eqref{sads:complexity} is constant in the time interval $-\tau_0\leq \tau\leq \tau_0$. Using \eqref{sads:tau0}, the constant value of the complexity is (for $n=2$)
\begin{equation}
	\label{sads:C.const}
	\mathcal{C}_0 = \frac{4L^2\alpha^2(1+\alpha^2)}{\pi\hbar G(1+3\alpha^2)}\left[\frac{1}{2}\ln\left(1+\alpha^{-2}\right)+\frac{2+3\alpha^2}{\sqrt{4+3\alpha^2}}\arctan\sqrt{3+4\alpha^{-2}}\right]~.
\end{equation}
If we perform an expansion in the large black hole limit, $\alpha\gg 1$, we find
\begin{align}
	\notag
	\mathcal{C}_0 &= \frac{4L^2}{3\sqrt{3}\hbar G}\left[\alpha^2 + \frac{\sqrt{3}}{\pi}+\frac{2}{3}-\frac{2}{27}\alpha^{-4} + O\left(\alpha^{-6}\right)\right]\\
	\label{sads:C.const2}
	&= \frac{4}{3\sqrt{3}\pi}S + \frac{4\pi^2C_T}{81}(9+2\sqrt{3}\pi) - \frac{8\pi^{11}C_T^3}{3^7\sqrt{3}}\frac{1}{S^2} + O\left(S^{-3}\right)~.
\end{align}
The final expression has been written in terms of physical expressions, in particular the black hole entropy \eqref{sads:entropy} and the central charge of the boundary CFT \cite{Buchel:2009sk}, $C_T = \frac{3L^2}{\pi^3\hbar G}$. We can interpret this constant value as the complexity of formation of the black hole from the empty AdS space-time\cite{Chapman:2016hwi}. The leading order of \eqref{sads:C.const2} shows the same linear relation between complexity of formation and black hole entropy highlighted in \cite{Chapman:2016hwi, Akhavan:2019zax}, at least up to a numerical factor. The sub-leading terms are different, and in particular we do not have the logarithmic divergence term that shows up in \cite{Chapman:2016hwi, Akhavan:2019zax}. These differences are due to the fact that a different parameterization was used in the computation. In \cite{Chapman:2016hwi} an affine parameterization is used, so that the boundary contribution of the null segments can be discarded. Incidentally, such a parameterization gives the same result as with the parameterization invariant action, which was used in \cite{Akhavan:2019zax}.

Last, for $\tau <\tau_0$, the complexity \emph{decreases} linearly. This behaviour does not appear to be physical, and we interpret it as an artifact of the eternal, two-sided black hole. In subsection~\ref{thermo} we will show that treating the black hole as a thermofield double state removes this artifact as well as the plateau at small $\tau$.

\subsection{Parameterization invariant action}
\label{sads:ct.ruin}

As we mentioned above, the complexity \eqref{sads:complexity} is not invariant under a change of parameterization of the null boundaries. Because parameterization independece may be considered as a necessary feature, we will calculate here the counter term \eqref{actionWdW:ct} that would render the action invariant. 

The quantity we need to compute the counter term is the null expansion along the boundaries
\begin{equation}
\label{sads:Theta}
	\Theta = n \frac{\rmd}{\rmd\lambda}\ln r(\lambda) = \frac{n}{r} k^r~.
\end{equation}

First, we consider the causal diamond of subsection~\ref{sads:caus.diam}. 
For the segment $N_1$ we readily find
\begin{align}
\notag
    \frac{I_{\mathrm{c.t.}, N_1}}{2\Omega_n} &= n\int\limits_{r_1}^{r_2}\rmd r\,r^{n-1}
    	\ln \frac{n\tilde{l}}{r} 
    	+ n \int\limits_{r_1}^{r_2}\rmd r\,r^{n-1}\sigma_1 \\
\label{sads:Ict1}
    	&= r_2\left(\ln\frac{n\tilde{l}}{r_2}+\frac{1}{n}\right)
    	 - r_1\left(\ln\frac{n\tilde{l}}{r_1}+\frac{1}{n}\right) 
    	 + n \int\limits_{r_1}^{r_2}\rmd r\,r^{n-1}\sigma_1~.
\end{align}
It is evident that the dependence on the parameterization function $\sigma_1$ in \eqref{sads:Ict1} cancels the corresponding integral term in \eqref{sads:IN1}. The contributions of the other boundaries take essentially the same form, and the total counter term reads
\begin{align}
\label{sads:Ict}
    \frac{I_{\mathrm{c.t.}}}{2\Omega_n} &= 
      2r_2\left(\ln\frac{n\tilde{l}}{r_2}+\frac{1}{n}\right)
    - 2r_1\left(\ln\frac{n\tilde{l}}{r_1}+\frac{1}{n}\right)
    - 2r_3\left(\ln\frac{n\tilde{l}}{r_3}+\frac{1}{n}\right)
    + 2r_4\left(\ln\frac{n\tilde{l}}{r_4}+\frac{1}{n}\right)\\
\notag
    &\quad + n\int\limits_{r_1}^{r_2}\rmd r\,r^{n-1}\sigma_1
	+n\int\limits_{r_3}^{r_2}\rmd r\,r^{n-1}\sigma_2 
	+n\int\limits_{r_3}^{r_4}\rmd r\,r^{n-1}\sigma_3 
	+n\int\limits_{r_1}^{r_4}\rmd r\,r^{n-1}\sigma_4~.
\end{align}
The action in the causal diamond, including the counter term, is obtained by adding \eqref{sads:Ict} to \eqref{sads:CD}, which gives 
\begin{equation}
    \label{sads:I.tot}
    \frac{I_\mathrm{tot}}{2\Omega_n} 
    = 2 n\int\limits_{r_1}^{r_2}\rmd r\,r^{n-1}\ln\frac{n\tilde{l}\sqrt{|f|}}{r} 
    + 2 n\int\limits_{r_3}^{r_4}\rmd r\,r^{n-1}\ln\frac{n\tilde{l}\sqrt{|f|}}{r}~.
\end{equation}
Here, we have chosen to leave the counter term in integral form. As expected, the result is independent of the parameterization of the null boundaries. However, there is no way to set this action generically to zero.

To see what are the consequences of this and the differences with the result \eqref{sads:complexity}, we  compute the action of the whole WdW patch with the contribution of the counter term.
Let us focus on the more interesting case $\tau>\tau_0$. The counter term contribution of the four null boundaries reads
\begin{align}
\label{sads:IctSigma}
	\frac{I_{\mathrm{c.t.}}}{2\Omega_n} = n\int\limits_0^R\rmd r\, r^{n-1}
		\left(2\ln\frac{n\tilde{l}}{r}+\sigma_1+\sigma_4\right)
		+n \int\limits_{r_m}^R\rmd r\, r^{n-1}
		\left(2\ln\frac{n\tilde{l}}{r}+\sigma_2+\sigma_3\right)~.
\end{align}
In computing \eqref{sads:I.WdW} we have set $a_v = 0$ and $a_u = 1$, which by \eqref{sads:choice} implies  $\sigma_1 = \sigma_3 = \ln|f|$ and $\sigma_2 = \sigma_4 = 0$. Using this parameterization in \eqref{sads:IctSigma} we find
\begin{align}
\notag 
	I_{\mathrm{c.t.}} &= 
		4 n \Omega_n\int\limits_0^R\rmd r\, r^{n-1}\ln\frac{n\tilde{l}\sqrt{|f|}}{r} 
		+ 4 n \Omega_n \int\limits_{r_m}^R\rmd r\, r^{n-1}\ln\frac{n\tilde{l}\sqrt{|f|}}{r}\\
\label{sads:wdw.Ict}
	&= I_{\mathrm{c.t.}}^0 
	- 4n\Omega_n\int\limits_0^{r_m}\rmd r\, r^{n-1}\ln\frac{n\tilde{l}\sqrt{|f|}}{r}~,
\end{align}
where we have defined the constant contribution 
\begin{equation}
	I_{\mathrm{c.t.}}^0 = 8n\Omega_n\int\limits_0^R\rmd r\, r^{n-1}\ln\frac{n\tilde{l}\sqrt{|f|}}{r}~.
\end{equation}
$I_{\mathrm{c.t.}}^0$ is just the counter term of the case $|\tau|<\tau_0$, when the WdW patch touches both singularities. We note that it diverges in the $R\to\infty$ limit. 

We can now sum \eqref{sads:wdw.Ict} and \eqref{sads:I.WdW} to find the action on the WdW patch with the counter term,
\begin{equation}
\label{sads:Iwdwct}
	I_{\text{WdW,tot}} = 4 n \Omega_n \left[ \omega^{n-1}(\tau+\tau_0) 
	- \int\limits_0^{r_m}\rmd r\, r^{n-1}\ln\frac{n\tilde{l}\sqrt{|f|}}{r} \right] 
	+ I_{\mathrm{c.t.}}^0~.
\end{equation}
The second term in the brackets, which stems from the counter term, is also time-dependent, because $\tau = - r^\ast(r_m)$. The time derivative of the action is
\begin{equation}
\label{sads:IWdWtder}
	\dot{I}_{\mathrm{WdW,tot}}= 4n\Omega_n\left[ \omega^{n-1} 
	+ r_m^{n-1}f(r_m)\ln\frac{n\tilde{l}\sqrt{|f(r_m)|}}{r_m} \right]~. 
\end{equation}
Let us briefly discuss the second term in the brackets. At late times, when $r_m\to r_H$, it vanishes, reproducing the known linear growth of the complexity. For $\tau$ just above $\tau_0$, \ie when the WdW patch detouches from the past singularity just after the time interval with constant action, it tends to $-\infty$. This behaviour was observed already in \cite{Carmi:2017jqz}, where it was proposed to smooth out the spike in the complexity by averaging over a time interval shorter than the thermal time.

\begin{figure}[th]
	\begin{center}
		\includegraphics[width=.60\textwidth,align=c]{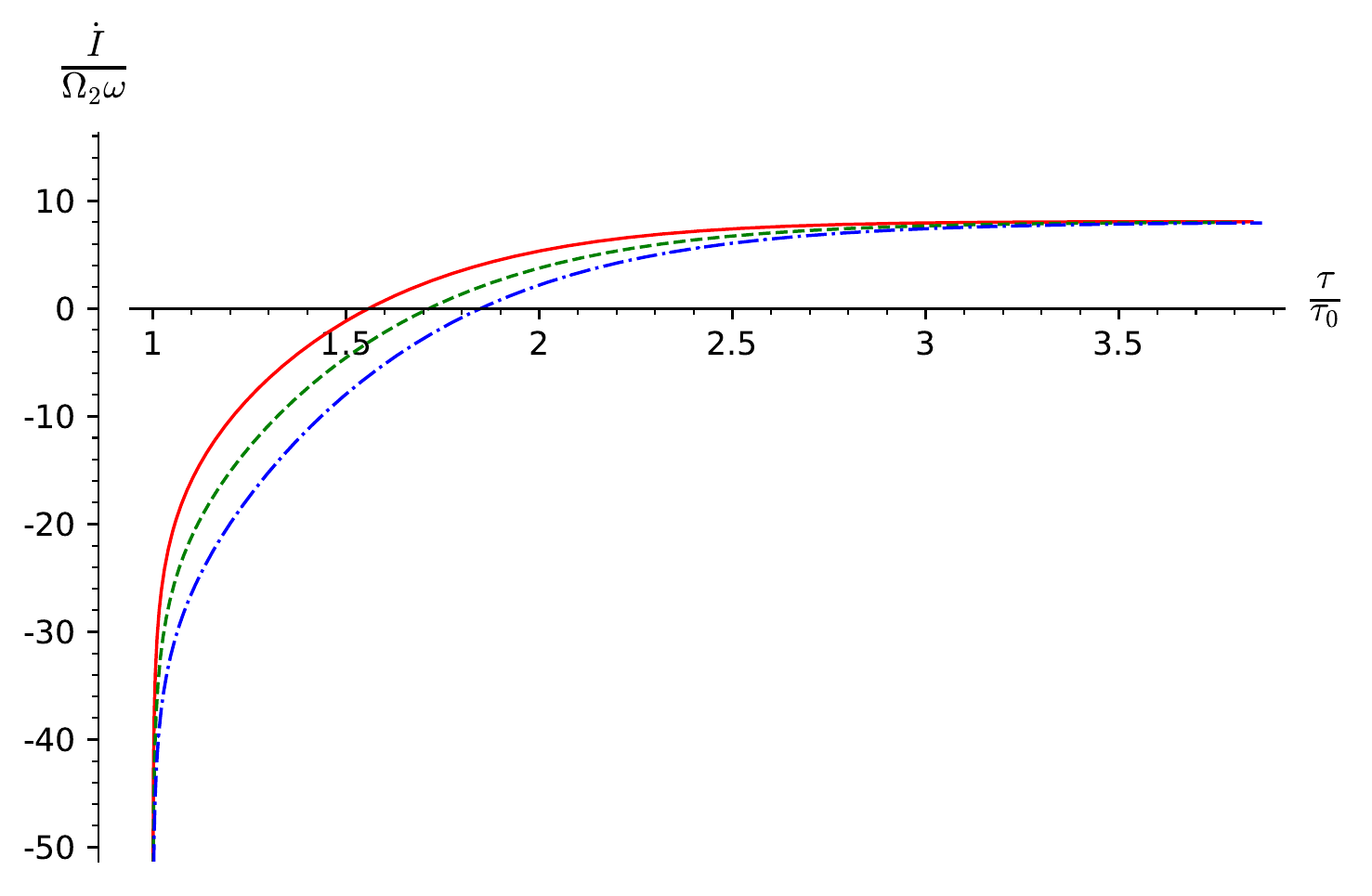}%
		\hfill%
		\caption{Time evolution of the action as a function of $r_m$ for different values of $\tilde{l}$. Solid line: $\tilde{l} = 0.5r_h$, dashed line: $\tilde{l} = r_h$, dash-dotted line: $\tilde{l} = 2r_h$. ($n=2$, $\alpha = 5$).\label{fig:plot}}
	\end{center}
\end{figure}

In Fig.~\ref{fig:plot} we plot \eqref{sads:IWdWtder} for some values of $\tilde{l}$ (we use the $n=2$ case for simplicity). We can see that at late times \eqref{sads:IWdWtder} agrees with the expected behaviour, namely linear growth, for any value of $\tilde{l}$. We also see however, that at early times the actions is decreasing. The time interval for which we have a decrease in the action is dependent on the arbitrary scale $\tilde{l}$.

The growth in complexity of the black hole is supposed to reflect the growth of the bulk volume behind the horizon, and as such it is expected to be at least non decreasing. Moreover, we find the fact that the time interval for which the action is decreasing depends on the arbitrary scale $\tilde{l}$ is even more unphysical.

\subsection{Thermofield double state}
\label{thermo}

As we have seen in the previous subsections, the action on a WdW patch in the eternal (two-sided) Schwarzschild-AdS black hole spacetime, with or without the counter term, is symmetric under time reflection, $\tau\to -\tau$. Moreover, it is constant in the time interval $|\tau| <\tau_0$. These two features are difficult to interpret when the action is taken as a measure for complexity. Now, we will show that taking literally the interpretation of the two-sided Schwarzschild-AdS black hole spacetime as the dual of a thermofield double state \cite{Maldacena:2001kr, Harlow:2014yka, ISRAEL1976107} resolves both issues.

Consider a quantum system with Hilbert space $\mathcal{H}$, Hamiltonian $H$ and the energy spectrum $H\ket{n}= E_n\ket{n}$. A thermofield double state is the following temperature-dependent entangled state in the doubled Hilbert space $\mathcal{H}\otimes \mathcal{H}$, 
\begin{equation}
\label{thermo:TFD}
	\ket{\psi} = \frac{1}{\sqrt{Z(\beta)}}\sum\limits_n\e{-\beta\frac{E_n}{2}}\ket{n_L}\otimes \ket{n_R}~.
\end{equation}
The subscripts $L$ and $R$ refer to the two copies of the Hilbert space, $Z(\beta)$ is the partition function of the theory, and $\beta$ the inverse temperature. The state \eqref{thermo:TFD} is prepared by an Euclidean path integral over the interval $\tau_E \in (0,\frac{\beta}{2})$, which prepares the state at some initial Lorentzian time $\tau=0$. For $\tau>0$, we will use the convention that $\tau = t_R = -t_L$, so that no time reflection is needed in the bulk description. This can be acieved by taking $H_L = -H_R$, or by considering a bra state $\bra{n_L}$ instead of $\ket{n_L}$. 

\begin{figure}[th]
	\begin{center}
		\includegraphics[width=.45\textwidth,align=c]{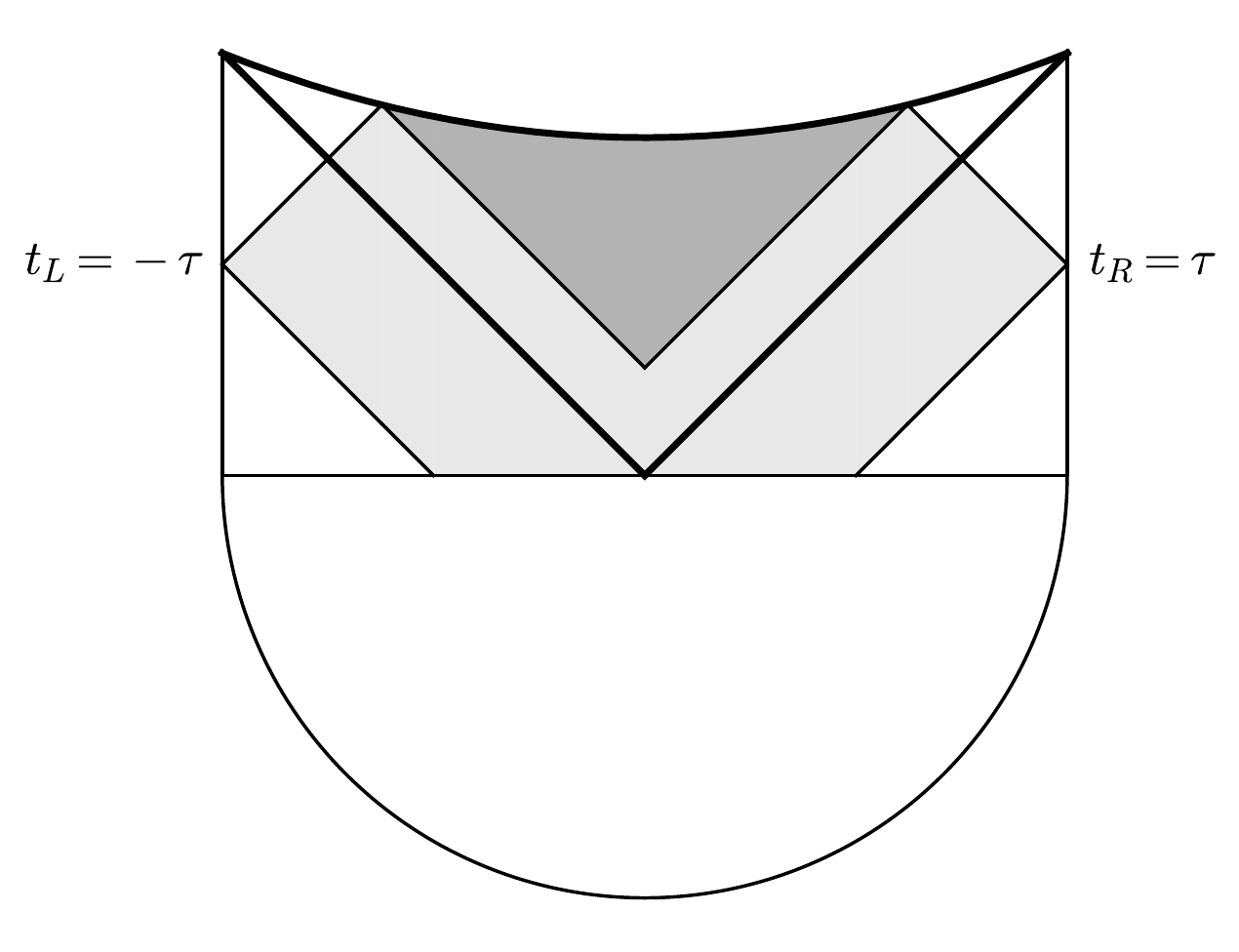}
		\caption{WdW patch in the thermofield double state. 
		The shaded area indicates a WdW patch, the darker part is the region that effectively gives the
		action. \label{fig:TFD.wdw}}
	\end{center}
\end{figure}

The bulk geometry of the Schwarzschild-AdS thermofield double state is obtained by gluing the half of the Lorentzian space-time with $\tau>0$ to the Euclidean geometry with $\tau_E \in (0,\frac{\beta}{2})$ \cite{Maldacena:2001kr}. Therefore, this construction gives us a precise definition of an initial time.
The resulting geometry is illustrated in Fig.~\ref{fig:TFD.wdw}, where we also show a WdW patch.

When we compute the action on the WdW patch using the prescription that causal diamonds do not contribute, then the region that effectively gives the action is the dark shaded area in Fig.~\ref{fig:TFD.wdw}, which borders with the singularity. We note that the triangular regions bordering with the $\tau=0$ spacelike hypersurface, which remain after removing causal diamonds, do not contribute. This is because they effectively add up to a causal diamond and the $\tau=0$ hypersurface has zero extrinsic curvature, so that the corresponding surface term vanishes. Therefore, the result is simply given by \eqref{sads:I.WdW},
\begin{equation}
\label{thermo:complexity}
	\mathcal{C} =  \frac{4M}{\pi \hbar}\left(\tau + \tau_0\right)~.
\end{equation} 
It is evident that the complexity growth is now linear at all times $\tau>0$.\\
When $\tau = 0$, the complexity is half the complexity of the double sided black hole, as we would expect since the WdW patch touches only one singularity. Consequently, the complexity of formation of the black hole is also halved. 

\begin{figure}[th]
	\begin{center}
		\includegraphics[width=.45\textwidth,align=c]{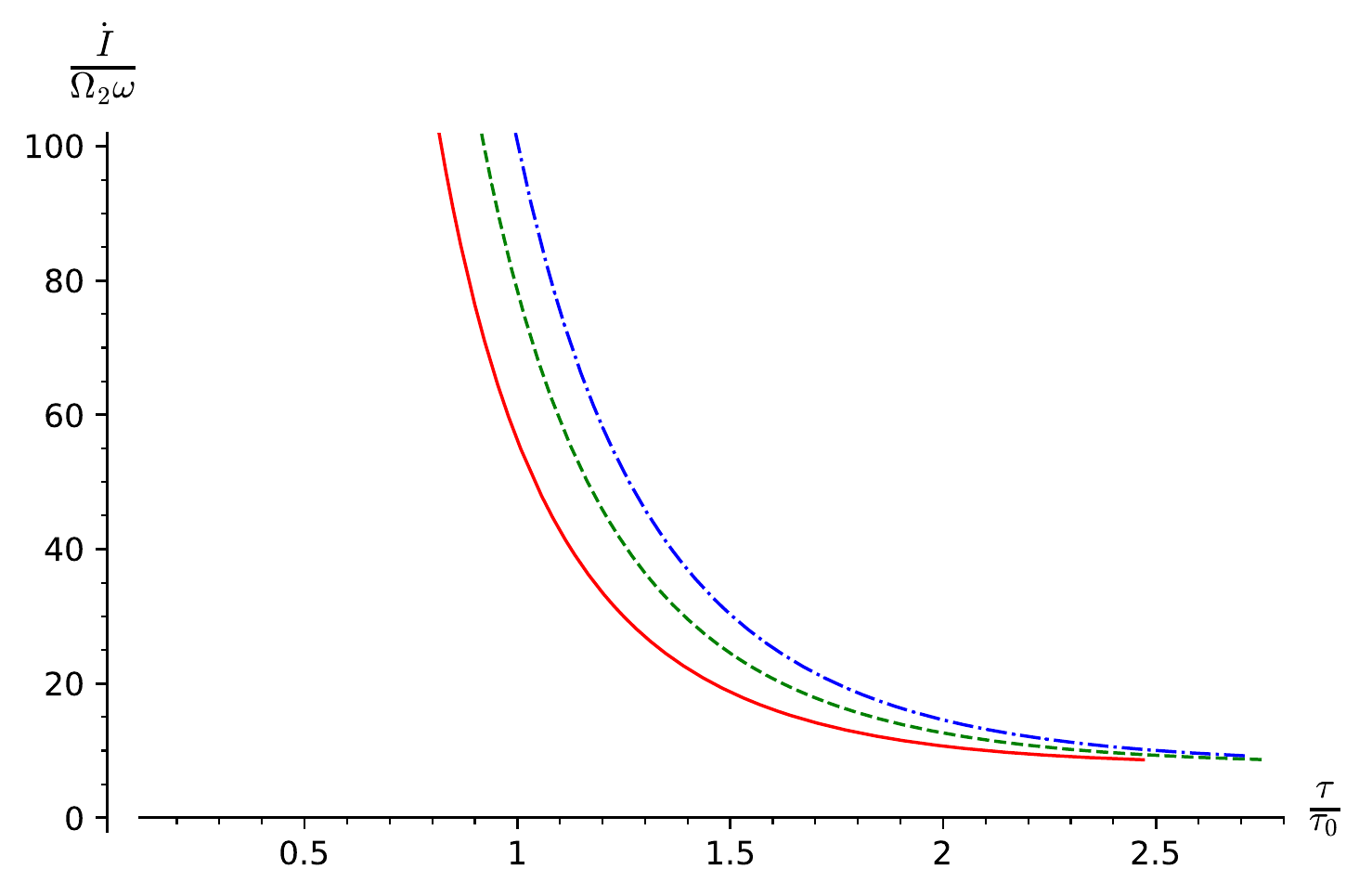}
		\caption{Time dervative of the action for different values of $\tilde{l}$. Solid line: $\tilde{l} = 0.5r_h$, dashed line: $\tilde{l} = r_h$, dash-dotted line: $\tilde{l} = 2r_h$. ($n = 2$, $\alpha = 5$).\label{fig:IdotTFD}}
	\end{center}
\end{figure}

We can quite easily see how the action would look like if we included again the counter term contribution. Such contribution takes the same form as \eqref{sads:wdw.Ict}, we only need to exchange $r_m$ with the radius $r_0$ in which the wdw patch intersects the boundary at $\tau=0$. The $\tau$-dependence of $r_0$ is given implicitly by: $r^\ast(r_0) = -\tau$. When $\tau = 0$ we have $r_0 = R$, and as $\tau$ grows to $\infty$ $r_0$ approaches $r_H$. The action with the counter term is then:
\begin{equation}
\label{thermo:Iwdwct}
	I_{\text{WdW,tot}} = 4 n \Omega_n \left[ \omega^{n-1}(\tau+\tau_0) 
	- \int\limits_0^{r_0}\rmd r\, r^{n-1}\ln\frac{n\tilde{l}\sqrt{|f|}}{r} \right] 
	+ I_{\mathrm{c.t.}}^0~.
\end{equation}
This expression suffers the same problems that affect \eqref{sads:Iwdwct}. The term $I_{\mathrm{c.t.}}^0$ introduces a divergence in the complexity, and the time derivative reads
\begin{equation}
	\label{thermo:IWdWtder}
	\dot{I}_{\mathrm{WdW,tot}}= 4n\Omega_n\left[ \omega^{n-1} 
	+ r_0^{n-1}f(r_0)\ln\frac{n\tilde{l}\sqrt{|f(r_0)|}}{r_0} \right]~. 
\end{equation}
Now $r_0$ goes from $R$ to $r_H$ as time increases. As show in Fig.\ref{fig:IdotTFD}, in the limit $R\to\infty$, when $\tau \to 0$, \eqref{thermo:IWdWtder} is divergent, while at late times it agrees again with the previous result, for any value of $\tilde{l}$. However as an improvement with respect to \eqref{sads:IWdWtder}, the action is now always increasing.
\section{BTZ black hole}
\label{btz}

In this section, we briefly discuss the special case of tridimensional bulk space ($n=1$). The vacuum solution, AdS$_3$, is not different from its higher dimensional counterparts, and the discussion of section~\ref{ads} is unchanged. 

The black hole solution is different instead. The BTZ black hole\cite{Banados:1992wn} is an asymptotically AdS$_3$ solution to gravity with a negative cosmological constant, and it is actually an orbifold of AdS$_3$ space-time. The solution is defined by the metric 
\begin{equation}
\label{btz:metric}
\rmd s^2 = -f(r) \rmd t^2 +f(r)^{-1} \rmd r^2 + r^2 \rmd\theta^2~,
\end{equation}
with 
\begin{equation}
\label{btz:f}
f(r) = \frac{r^2}{L^2} - m~.
\end{equation}
The black hole has an horizon at $r_H = \sqrt{m}L$ and a singularity at $r=0$.\footnote{We only consider $m>0$. For $m<0$, the solution represents a point-like conical defect without horizon.} The mass, temperature, and entropy of the black hole read
\begin{equation}
M = \frac{m}{8G},\quad T = \frac{r_H}{2\pi L^2},\quad S = \frac{\pi r_H}{2G},
\end{equation}
respectively. The causal structure and the Penrose diagram of the BTZ black hole are identical to the Schwarzschild case shown in Fig.\ref{fig:schwads}. The tortoise radial coordinate can be computed explicitly 
\begin{equation}
\label{btz:tortoise}
r^\ast(r) = \int\limits_{\infty}^r \frac{\rmd r}{f(r)} = \frac{r_H}{2m}\ln\frac{|r-r_H|}{r+r_H}~.
\end{equation} 
The WdW-patch is similar to the patch of the Schwarzschild case (see Fig.\ref{fig:wdw}), with only one difference. Since $t_0 = -r^\ast(0) = 0$, the patch is in contact with only one of the two singularities when $|\tau|>0$, but touches both the future and past singularities when $\tau = 0$.  

The computation of the action of the WdW-patch proceeds in the same way we have already shown for the AdS-Schwarzschild black hole with no significant differences. In particular, the action in any causal diamond is vanishing when choosing a parameterization like \eqref{sads:choice}, with $f(r(\lambda))$ given by \eqref{btz:f}. The contribution to the action of the WdW-patch can then only come from the regions in contact with the singularity (see Fig.\ref{fig:wdw_dark}), and the complexity is 
\begin{equation}
	\label{btz:C}
	\mathcal{C} = \frac{4M}{\pi\hbar}|\tau|~.
\end{equation} 
The fundamental difference respect to the complexity of the AdS-Schwarzschild black hole is that the complexity is linearly increasing for all $\tau>0$, with no initial constant plateau. In particular, we observe that the complexity is zero when $\tau = 0$, meaning that the complexity of formation of the BTZ black hole with respect to empty AdS space-time is zero. In \cite{Chapman:2016hwi}, the complexity of formation of the BTZ black hole was computed by subtracting the complexity of AdS$_3$ from the complexity of the black hole, using an affine parameterization to compute the null boundaries' contributions to the action. There it was found that the complexity of formation depends on the type of vacuum chosen for the boundary theory, and it is zero only in the case of the Ramond vacuum. This suggests that the criterion proposed for choosing a parameterization, namely that the action of any empty causal diamond must vanish, automatically chooses the type of vacuum of the boundary theory.   
\section{Conclusions}
\label{conc}

In this paper, we have reconsidered the holographic complexity of pure AdS space-time and the AdS-Schwarzschild black hole in the $C=A$ approach. The novelty of our treatment lies in the departure from the requirement that the on-shell action be invariant under reparameterizations of the null components of the boundary of the WdW patch. This requirement would mandate a counter term in addition to the minimal action determined by the variational principle and would leave the renormalization scale $\tilde{l}$ as the only parameter. 
Instead, we regard the parameterization dependence as a feature that allows to describe physically different situations. On the bulk side, the action terms on the null boundaries describe the heat content on these boundaries \cite{Chakraborty:2019doh}. We think that the interpretation in the dual CFT is related to the details of the definition of circuit complexity, \ie the reference state and the set of elementary gates, but we do not have anything precise to say on this point. 
Having disposed of reparameterization invariance, we have introduced a new criterion that selects physically sensible parameterizations. Our criterion is that \emph{the action in any static vacuum causal diamond vanishes}.
This immediately sets the action on the WdW patch in pure global AdS space-time to zero, fixing this state as the reference state. This was not possible in approaches similar to holographic renormalization.
For the AdS-Schwarzschild and BTZ black holes, our criterion renders the action intrinsically finite, because the regions near the space-time boundary can now be discarded. In these cases, the region that effectively gives the complexity lies entirely behind the horizon and borders with the singularity. The new criterion entails that the preferred parameterization functions on the null boundaries are given in terms of the logarithm of the blackening function, see \eqref{sads:choice}, up to free numerical coefficients that must add up to unity between two neighbouring boundaries.

In the case of the eternal (two-sided) black hole, our calculation not only confirms the linear growth of complexity at late times, but makes the growth linear at all times $\tau>\tau_0$, where $\tau_0$ is a critical time, saturating the Lloyd bound. In particular, the peculiar dip immediately after the WdW patch detaches from the past singularity \cite{Carmi:2017jqz}, which was found in the counter term approach, is absent. For $0<\tau<\tau_0$, the complexity is constant. This constant value is interpreted as a complexity of formation and agrees up to a numerical factor and to leading order with previous results for that quantity \cite{Chapman:2016hwi, Akhavan:2019zax}. 
Only the tridimensional case is special. For the BTZ black hole, we find linear growth for all $\tau>0$, with $\tau_0=0$. We also considered the thermofield double state interpretation of the AdS-Schwarzschild black hole replacing the unphysical region of negative $\tau$ (where complexity decreases) with an appropriate Euclidean space-time region that creates the thermofield double state. In this case, the complexity turns out to be linear at all times $\tau>0$, starting at a positive value that can again be interpreted as a complexity of formation. 

One might object that the divergence of holographic complexity in the reparameterization-invariant approach, when the cut-off is sent to infinity, reflects the generic divergence of circuit complexity when the precision parameter $\epsilon$ is sent to zero. Although it is true, this argument has a loop hole. Physically, one can reformulate this expectation by saying that constructing, with \emph{infinite} precision, a generic target state from a given reference state using a given set of elementary operations requires, generically, infinitely many operations. However, this is only the generic situation, and holographic complexity still is divergent when generic parameterizations of the null boundaries are considered. It all depends on what happens to the set of elementary operations as the limit $\epsilon\to 0$ is taken. In the extreme case in which \emph{all} unitary operations are allowed as elementary operations, only a single operation would be needed to obtain the target state, trivializing complexity. Therefore, one should think that for parameterizations satisfying our criterion the set of elementary operations is suitably enlarged in the limit $\epsilon\to 0$, such that the complexity remains finite.\footnote{There is an easy, although not very rigorous, analogy in binary terms. As $\epsilon$ decreases, the number of bits necessary to distinguish a positive number from zero increases. At the same time, the length of binary numbers manipulated by the elementary operations should also increase.} 

An obvious question that arises is how to apply our new criterion to charged or rotating black holes. In such black holes, which typically have an inner and an outer horizon, the WdW patches are always causal diamonds, which never come in contact with the singularity. Asking that the action vanish in any causal diamond would then imply that the holographic complexity vanishes, which cannot be the correct answer. Therefore, rather than applying the criterion to any causal diamond, one should extend the parameterization functions found here to the more general cases. In particular, we found here that the preferred choice of the parameterization functions is in terms of the blackening function, see \eqref{sads:choice}. This choice of parameterization must be extended by analogy to the more general cases and gives rise to a non-vanishing action of causal diamonds in the charged and rotating cases, simply because they are not vacuum or static. 
We have considered these cases, as well as Vaidya space-time, in a follow-up paper \cite{Mounim:2021bbb}. 

Another interesting direction would be the study of the complexity behaviour under conformal transformations.\cite{Flory:2019kah, Flory:2019dqx}, since when using the "Complexity=Action" conjecture, the counter term gives rise to effects of difficult physical interpretation\footnote{We thank Mario Flory for pointing this out to us.}

\section*{Acknowledgements}
This work was supported partly by the INFN, research initiative STEFI.
	
\bibliographystyle{JHEPnotes}
\bibliography{adscft,nsf,complexity}
\end{document}